\documentclass[11pt]{article}
\pdfoutput=1

\usepackage{amssymb}
\usepackage{amsmath}
\usepackage{epsfig}

\begin{document}
\newtheorem{theorem}{Theorem}
\newtheorem{corollary}{Corollary}
\newtheorem{definition}{Definition}
\newtheorem{lemma}{Lemma}

\newcommand{\define}{\stackrel{\triangle}{=}}
 
\pagestyle{empty}

\def\QED{\mbox{\rule[0pt]{1.5ex}{1.5ex}}}
\def\proof{\noindent\hspace{2em}{\it Proof: }}

\date{}
\title{Exploiting Channel Correlations  - Simple Interference Alignment Schemes with no CSIT} 
\author{\normalsize  Syed A. Jafar\footnote{This work was supported by DARPA ITMANET under grant UTA06-793,  by NSF under grants 0546860 and 
CCF-0830809 and by ONR YIP under grant N00014-08-1-0872.}\\
University of California Irvine\\
      {\small \it E-mail~:~\{syed\}@uci.edu} \\
       }
%% Notes
\maketitle

\thispagestyle{empty}

%%%%%%%%%%%%%%%%%%%% Abstract %%%%%%%%%%%%%%%%%%%%%%%%%
\begin{abstract}
We explore 5  network communication problems where the possibility of interference alignment, and consequently the total number of degrees of freedom (DoF) with channel uncertainty at the transmitters are unknown. These problems share the common property that in each case the best known outer bounds are essentially robust to channel uncertainty and represent the outcome with interference alignment, but the best inner bounds -- in some cases conjectured to be optimal --   predict a total collapse of DoF, thus indicating the infeasibility of interference alignment under channel uncertainty at transmitters. Our main contribution is to show that even with no knowledge of channel coefficient values at the transmitters, the knowledge of the channels' correlation structure can be exploited to achieve interference alignment. In each case, we show that under a staggered block fading model, the transmitters are able to align interference without the knowledge of channel coefficient values. The alignment schemes are based on linear beamforming -- which can be seen as a repetition code over a small number of symbols --  and involve delays of only a few coherence intervals.
\end{abstract}
\newpage
%%%%%%%%%%%%%%%%%%%% Introduction %%%%%%%%%%%%%%%%%%%%%%%%%
\section{Introduction}
There is much recent research activity aimed at characterizing the degrees of freedom of wireless networks. The topic is of interest due to the novel insights -- in particular those related to interference alignment --  that have emerged out of this perspective. Interference alignment refers to the design of signals in such a way that they cast overlapping shadows at the receivers where they constitute interference while remaining separable from the interference at the receivers where they are desired. Interference alignment schemes have been found for a variety of networks including  X networks \cite{MMK,Jafar_Shamai}, compound broadcast channel \cite{Weingarten_Shamai_Kramer, Gou_Jafar_Wang}, interference networks \cite{Cadambe_Jafar_int,Bresler_Parekh_Tse,Cadambe_Jafar_Shamai,Gou_Jafar, Sridharan_Jafarian_Vishwanath_Jafar, Etkin_Ordentlich, Nazer_Gastpar_Jafar_Vishwanath, Motahari_Gharan_Khandani},  cellular networks \cite{Suh_Tse, Cadambe_Jafar_Wang},  multihop (relay) networks \cite{Jeon_Chung, Jeon_Chung_Jafar},  and bidirectional relay networks \cite{Lee_Lim}. These include linear alignment schemes over signal spaces (introduced in \cite{Jafar_Shamai, Weingarten_Shamai_Kramer}), signal level and lattice alignment schemes (introduced in \cite{Bresler_Parekh_Tse,Sridharan_Jafarian_Vishwanath_Jafar}), asymptotic alignment schemes over a large number of dimensions (introduced in \cite{Cadambe_Jafar_int}), propagation delay based alignment \cite{AsilomarDelay,Grokop_Tse_Yates},  asymmetric complex signaling schemes (introduced in \cite{Cadambe_Jafar_Wang}), rational/irrational scaled lattice alignment schemes (introduced in \cite{Etkin_Ordentlich,Motahari_Gharan_Khandani}) and ergodic alignment schemes (introduced in \cite{Nazer_Gastpar_Jafar_Vishwanath}). 

While the capacity benefits of interference alignment schemes have been shown to be substantial, the caveat behind most of these results has been the assumption of perfect,  and sometimes global, channel state information at the transmitters (CSIT). Indeed, in the absence of channel knowledge it is well known that the degrees of freedom of many networks collapse entirely to what is achievable simply by orthogonal time-division among users \cite{Caire_Shamai, Jafar_mobile}. With partial channel knowledge the possibility of interference alignment becomes very intriguing and very little is known about the network degrees of freedom. A common trend is discernible in the diverging inner and outer bounds for various networks under partial CSIT \cite{Lapidoth_Shamai_Wigger_BC,Weingarten_Shamai_Kramer,Huang_Jafar_Shamai_Vishwanath}. While the best known inner bounds predict a total collapse of degrees of freedom, the best known outer bounds remain fairly robust to channel uncertainty. A few representative examples  extracted from prior work, that illustrate this trend,  are listed below.
\begin{enumerate}
\item {\bf MISO BC with no CSIT for one User: } Consider the multiple input single output (MISO) broadcast channel (BC) with $2$ antennas at the transmitter and $2$ receivers (users), each equipped with a single antenna. The channel state of one user is known perfectly to the transmitter while the other users' channel state is unknown. Assuming that the channel states are generic and fixed throughout the duration of communication, the best known outer bound on the total DoF, obtained in \cite{Weingarten_Shamai_Kramer}, is equal to $\frac{3}{2}$. However, the best known inner bound on the total DoF if user 2's channel state is drawn from some continuous distribution, is only $1$ - which is trivially achieved by orthogonal time-division between users. The inner bound is conjectured to be optimal.
\item {\bf MISO BC with no CSIT for both Users: }For the same 2 user MISO BC setting, if the channel states of both users are unknown to the transmitter, and held fixed, then the best known outer bound on the total DoF, also found in \cite{Weingarten_Shamai_Kramer} is $\frac{4}{3}$. If the channel states are time-varying, the best known outerbound on the total DoF, derived in \cite{Lapidoth_Shamai_Wigger_BC} is still equal to $\frac{4}{3}$. In both cases the best inner bound is only $1$. In both cases, the inner bound is conjectured to be tight.
\item {\bf X Channel: }The best known outer bound on the total DoF of the X channel with 2 transmitters and 2 receivers -- whether the channel is time-varing or held constant-- follows from the arguments as presented in \cite{Jafar_Shamai} and is equal to $\frac{4}{3}$ -- same as with perfect channel knowledge. The best known inner bound  in this case is also only 1.
\item {\bf MIMO Interference Channel: } A similar open problem is pointed out in the context of the 2 user time-varying MIMO interference channel studied in \cite{Huang_Jafar_Shamai_Vishwanath}. Consider the setting with $1,3$ antennas at transmitters $1,2$ and $2,4$ antennas at receivers $1,2$ respectively. If user 1 achieves 1 DoF (his maximum possible DoF), then what is the maximum DoF simultaneously achievable by user 2? The outer bound for the DoF achieved by user 2 found in \cite{Huang_Jafar_Shamai_Vishwanath} is $\frac{3}{2}$ but the best known achievable DoF for user 2 is only equal to $1$, which is achieved by simple zero forcing at the receivers  and no interference alignment.
\item {\bf Interference Network: } The best known outer bound on the total DoF of the interference channel with $K$ transmitters and $K$ receivers -- whether the channel is time-varing or held constant-- follows from the arguments as presented in \cite{Cadambe_Jafar_int} and is equal to $\frac{K}{2}$ -- same as with perfect channel knowledge. The best known inner bound with channel uncertainty in this case is also only 1.
\end{enumerate}

 Note that in all these cases the channel coefficients are assumed to be generic -- i.e. drawn from a continuous distribution-- and their values are assumed known to the receivers perfectly. While this may include some degenerate cases where indeed the collapse of DoF is trivially evident due to e.g. rank deficiency of channel coefficients matrix, these degenerate cases correspond to a set of measure 0. The DoF inner and outer bounds mentioned above are meant in the "almost surely" sense, i.e. for almost all values of channel coefficients. 

If the channel uncertainty at the transmitters is not spread over a continuum, especially if the channel coefficient values are drawn from a set of finite cardinality, the results can be quite different from the pessimistic conjectures that favor the inner bounds. This setting is called the finite state compound network setting in \cite{Gou_Jafar_Wang}. It was originally thought that even in this setting the DoF will collapse to unity as the number of possible states increases \cite{Weingarten_Shamai_Kramer}. This conjecture was, however, disproven in \cite{Gou_Jafar_Wang}. In most cases listed above (including some generalizations to multiple users or antennas) it was found that the outer bounds are tight in the finite state compound network setting. While this is a positive result, it should be noted that the achievable schemes rely strongly on unbounded bandwidth expansion or unlimited resolution of channel coefficients e.g. through alignment of lattices scaled by transcendental numbers whose rational independence allows these superimposed lattices to remain separable in the high SNR regime \cite{Gou_Jafar_Wang, Motahari_Gharan_Khandani}. While the alignment problem becomes increasingly difficult and the achievable rates should degrade significantly as the number of states increases, the degrees of freedom metric appears to be too coarse to capture this penalty. Thus, while the degrees of freedom are found to be robust in the finite state compound setting, the alignment schemes themselves are quite fragile. Arguably, more than the robustness of interference alignment these results show the limitations of the finite state compound model, which does not capture channel uncertainty as well as previously believed, especially in the asymptotic high SNR setting.

This paper is motivated by the need to find \emph{robust} alignment schemes with the understanding  that {an interference alignment scheme is robust if the alignment is achieved even with the transmitters remaining oblivious of the values of channel coefficients which may be drawn from a continuum.}  The main contribution of the paper is summarized in the following new insight.

{\bf Key New Insight -}{\it Even if the transmitters have no knowledge of channel coefficient values, they can still align interference based on the knowledge of only the temporal correlation structures of the channel variations associated with different users.}

To better appreciate this result,  let us briefly review the conventional wisdom that has guided the choice of temporal correlation models in theory. Since wireless channels are time-varying, channel uncertainty is a problem for both transmitters and receivers. Due to the need for transmitters to use the channel knowledge in real-time before it gets old -- and therefore useless -- and also because of the feedback overhead needed to convey the receivers' channel estimates back to the transmitters as quickly as possible, CSIT is considered very challenging to obtain. CSIR on the other hand is more naturally available as the receiver observes the signals transmitted through the channel and can afford to wait until the end of transmission to refine its channel estimates. Clearly the quality of CSIR fundamentally depends on the rate of temporal variations of the channel coefficients -- the coherence time. Thus, research aimed at exploring channel uncertainty at receivers invariably takes into account the temporal correlations of channel fading. A common model for this is the block fading model \cite{wc_book_tse}, but other temporal correlation models, such as autoregressive processes \cite{Baddour_Beaulieu}, have been used in this context as well. In all cases, if the channel varies slowly enough, i.e. if the channel coherence time is large enough, the channel uncertainty at the receiver is found to be negligible as the cost of learning the channel is amortized over long coherence periods. For most common communication scenarios the coherence times are large enough that the theoretical assumption of perfect CSIR is considered reasonable. Even with perfect CSIR, if the goal is to study the impact of partial CSIT acquired through imperfect -- e.g. delayed -- feedback, then the temporal correlation model is crucial. However, if there is no feedback mechanism in place for acquiring CSIT, then the temporal correlation model is most commonly ignored. Hence the prevalence of the i.i.d. Rayleigh fading model -- where the channels take independent values every time slot --  in conjunction with perfect CSIR and no CSIT scenarios. The understanding is that the temporal correlations are crucial primarily for determining the quality of CSIR and the quality of CSIT acquired through feedback. CSIR is not an issue once perfect CSIR is assumed, and feedback is not an issue if the transmitter does not attempt to learn the channel coefficients by design. Therefore in this setting temporal correlations are largely inconsequential and in the interest of simplicity and tractability, they are best ignored. For point to point channels,  this understanding can be formalized through capacity results. Indeed, for a point to point wireless channel the ergodic capacity with perfect CSIR and no CSIT does not depend upon temporal correlations. So for example, the i.i.d. Rayleigh fading model with perfect CSIR and no CSIT has the same capacity whether the channels follow a block fading model (with i.i.d. variations across blocks) or vary independently from one symbol to next (coherence time 1).

In contrast, what the result of this work shows is that the conventional wisdom summarized above does not carry over to networks. Even with perfect CSIR and no CSIT, temporal correlations play a very important role in a network -- especially in terms of how the correlation structure varies from user to user -- because the difference in temporal correlation structures of different users creates opportunities for interference alignment. For example, consider a MISO BC where all channels are i.i.d. Rayleigh fading across users and antennas, with perfect CSIR and no CSIT. On the one extreme we have the case where all users follow the same temporal correlation model. In this case all receivers are statistically equivalent,  and it is well known that the DoF of this channel collapses to unity \cite{Caire_Shamai}. In fact time division is capacity optimal  at any SNR and the capacity does not depend on the correlation structure at all. Now, contrast this with a setting where the users have different temporal correlation structures, e.g. block fading with different coherence times. In this paper we find that based on just the knowledge of the temporal correlation structure, \emph{without any knowledge of the values of the channel coefficients}, the transmitters are able to design their signals such that interference is aligned, so much so that even the outer bounds considered unachievable in the benchmark scenarios outlined above, are achievable. Moreover, the coding scheme proposed in this paper to achieve this alignment with no CSIT has several desirable features as highlighted below.
\begin{itemize}
\item No knowledge of the values of channel coefficients is required at the transmitters. This in contrast to the infinite precision channel knowledge needed at transmitters to achieve the DoF outer bounds, for almost all interference alignment schemes proposed so far.
\item The encoding is a simple linear beamforming scheme. This is in contrast to sophisticated lattice alignment schemes used in \cite{Bresler_Parekh_Tse, Sridharan_Jafarian_Vishwanath_Jafar, Etkin_Ordentlich, Motahari_Gharan_Khandani, Gou_Jafar_Wang}.
\item The size of supersymbols and the number of independently encoded streams over which beamforming takes place is small. This is in contrast to the long symbol extensions needed for the alignment scheme in \cite{Cadambe_Jafar_int}.
\item Alignment is achieved over very few channel coherence times. This is in contrast to the very long waiting times required for pairing complementary matrices in the ergodic interference alignment scheme of \cite{Nazer_Gastpar_Jafar_Vishwanath}.
\item The alignment scheme works equally well in both the complex and real setting. This is in contrast to the DoF results for the finite state compound channel where it has not been possible so far to extend the results found for the real setting to the complex setting in general \cite{Gou_Jafar_Wang}.
\end{itemize}

On the one hand the alignment schemes introduced in this work satisfy the robustness requirement of needing no knowledge of the values of the channel coefficients. On the other hand, the DoF results obtained in this work rely strongly on the block fading model. Clearly, the model itself is an idealization because in practice the channel states are continuously varying and will not stay \emph{exactly} the same even across consecutive symbols. In this light, there are two interpretations of the results presented here. The pessimistic interpretation is that it reveals the limitations of the block fading model for networks. Indeed the main difference between our present model which allows optimistic results, and the model considered by Lapidoth et. al. in \cite{Lapidoth_Shamai_Wigger_BC} to arrive at their pessimistic conjecture outlined above, is that we allow channel coefficients to remain unchanged over a coherence interval. If the coefficients are allowed to continuously vary, then indeed we do not rule out the possibility that the DoF may collapse as suggested by the Lapidoth-Shamai-Wigger conjecture \cite{Lapidoth_Shamai_Wigger_BC}.

However, an optimistic interpretation would take a different view. Ultimately what matters is the implication of these results for practical systems, which never operate at  \emph{infinite} SNR. A high SNR result is mainly interesting (from a practical perspective) for the insights it provides about the \emph{finite} SNR scenario. At finite SNR, however,  we note that the linear interference alignment schemes outlined here allow a graceful degradation as the channel model transitions from block fading to continuously but slowly varying -- mainly because the resulting SINRs are continuous functions of the channel perturbations. Thus, even if the Lapidoth-Shamai-Wigger conjecture is correct and the DoF under continuous variations collapse to unity, this extremely negative outcome may only be an artifact of the infinite SNR regime. Following this line of thought to its logical conclusion, the overarching insight is that some idealization of channel models -- e.g. block fading -- may be necessary to find \emph{useful} results in the infinite SNR regime just to avoid the artifacts of the infinite SNR regime themselves dominating the conclusions. These ideal channel models and infinite SNR results may still provide useful insights in practice, because as we move to more realistic channel models, we also move away from the infinite SNR regime. These two perturbations work in opposite directions and hence could be expected to cancel each other. For example, we seek perfect interference alignment and separation of interference from signal spaces for DoF characterizations. With realistic channel uncertainty models, perfect interference alignment may not be achievable, but it is also important to note that in the finite SNR setting perfect interference alignment is also not desirable.  The best solutions will also favor the desired signal power even at the cost of admitting some interference. In this light the conclusions of the present paper, as well as most DoF results, should be seen as only providing insights into the possibilities that may be exploited in the finite SNR regime --- and not as comprehensive solutions to the practical problem itself.

\section{Channel Model - Staggered Block Fading}
We will assume the block fading model, where the channel states are constant for a duration equal to the channel coherence time $T$ and then switch to a different generic (i.e. drawn from a continuous distribution but not necessarily independent) value. Unless explicitly mentioned otherwise, the channel states are assumed to be known perfectly to the receivers and not known to the transmitters. Since we are interested in interference alignment, it is important to recall that the enabling premise of interference alignment is the \emph{relativity of alignment}, i.e., receivers should be statistically distinguishable as seen from the transmitters. Otherwise, if the receivers are statistically equivalent it is easy to show that the DoF do collapse to unity (assuming single antenna receivers) as all messages can be decoded by the same receiver \cite{Caire_Shamai, Jafar_mobile}. In this paper we will assume this relativity of alignment comes from the fact that the channel coherence times vary across users and their fading blocks are not synchronized. In other words, the temporal boundaries of the fading block for each channel may be different. We refer to this assumption as \emph{staggered block fading}. 

{\it Remark: } The staggered block fading model should not be confused with the staggered propagation delay model in \cite{AsilomarDelay} where also interference alignment is achieved without knowledge of channel coefficient values. No special assumptions about propagation delays are involved here and it bears repeating that the staggered blocks only refer to the temporal correlation structure of channel variations across various users.

Before proceeding further, we briefly discuss the physical justification for the staggered block fading model. From the perspective of achievable schemes used in this work, it is important to mention that the staggered block fading model can arise out of the difference in the channel coherence times for different users. Indeed, in a geographically distributed network,  users in more mobile local environments may have smaller channel coherence times than other users who are in relatively stationary local environments. Through an interleaving of symbols this difference in coherence times may be converted into the supersymbol structures needed in this paper. 
Further, while we will assume small values of coherence time interval (we assume $T=2$) throughout this paper, this is not a fundamental limitation for the schemes introduced here, as explained next.

\begin{figure}[!t]
\centering
\includegraphics[width=6in]{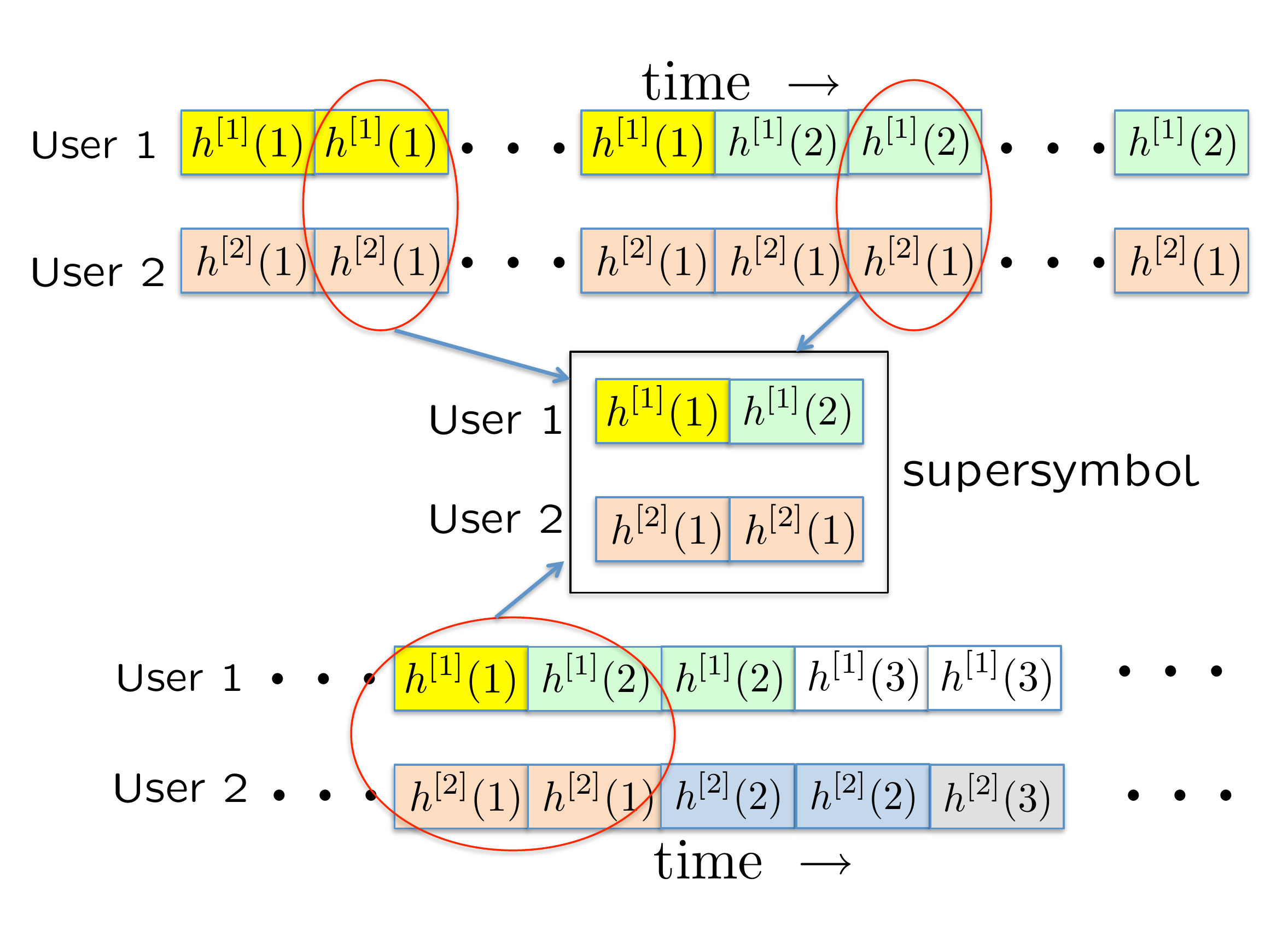}
\caption{Staggered Block Fading Model and the Supersymbol Structure}
\label{fig:staggerfade}
\end{figure}

Figure \ref{fig:staggerfade} illustrates this point. At the bottom of the figure is a staggered block fading model for two users with hypothetical channel state variables $h^{[k]}(n)$ where $k=1,2$ indicates the user and $n$ indicates the coherent block index. The coherence time $T=2$ for both users. While each user's channel stays fixed for 2 time slots, the boundaries of these fading blocks are staggered. If we take a snapshot over two symbols (called a supersymbol in the figure) we note that user 1's channel changes state, while user 2's channel stays fixed during this supersymbol. The same supersymbol can be seen to arise out of a more physically meaningful setting shown at the top of the figure. Suppose the users 1,2 have different coherent times $T_1, T_2$ (which can take large values), respectively, with $T_2=2T_1$, i.e. user 2's channel stays constant twice as long as user 1's channel state. Other than necessitated by this difference in channel coherence times, the fade blocks need not be staggered. However, by combining two symbols that correspond to the same coherent block of user 1 and two different coherent blocks of user 2, one can again create the same supersymbol as shown in the figure. Even if the $T_2$ is not exactly twice $T_1$, as long as $T_2\gg T_1$, similar pairings can be found to obtain the supersymbol structure shown in the center of the figure. Thus, the block fading model at the top which is easily physically motivated and the staggered block fading model at the bottom which may seem less natural, both give rise to the same supersymbol structure. For the purpose of the interference alignment schemes in this paper, the key is only the supersymbol structure. Thus, both block fading models are equivalent from that perspective. 

{\it Remark:} We assume throughout that the channel temporal correlation structure of all users is known perfectly to the transmitters. We still describe this as the ``no CSIT" setting, because the correlation structure only constitutes knowledge of channel statistics and not the specific realizations of the channel coefficients. However, it must be noted that this knowledge of the channel structure is essential to the interference alignment schemes proposed in this work. Clearly, learning the channel statistics at the transmitter represents much less overhead than learning the precise values of the channel coefficients.

Our goal in the next few sections is to consider each of the representative problems highlighted in the introduction and to show, in each case, that the outer bound is achievable with only the knowledge of the channel coherence structure at the transmitters.  Note that because our staggered block fading channel model is different from the context where the outer bounds are originally derived,  a formal derivation of the outer bounds for our channel model is needed. Some of the outer bounds follow in a trivial manner from previously known results. For example, the X channel outer bound is trivial - it is the same outer bound as with perfect CSIT. The same would be true for interference networks as well. However,  for the results concerning the MISO BC model, and for the MIMO interference channel problem,  a formal derivation of the outer bound under our precise channel model is required.  Since the focus of this work is on achievability of interference alignment, and we expect the outer bounds -- as well as the main derivation principles --  to remain unchanged (as already evident for the MISO BC  in the two extremes of the time-varying model \cite{Lapidoth_Shamai_Wigger_BC} and the fixed state compound setting \cite{Weingarten_Shamai_Kramer}), we postpone the formal proofs of these outer bounds to the full paper. Here we focus only on the achievability arguments. 

We expect the reader to be very familiar with the Shannon-theoretic definitions of achievable rates, capacity, degrees of freedom etc. which are used in this work only in the standard sense. Therefore, we will proceed directly to the technical content. 

\section{MISO BC with no CSIT for One User}
\begin{figure}[!t]
\centering
\includegraphics[width=4in]{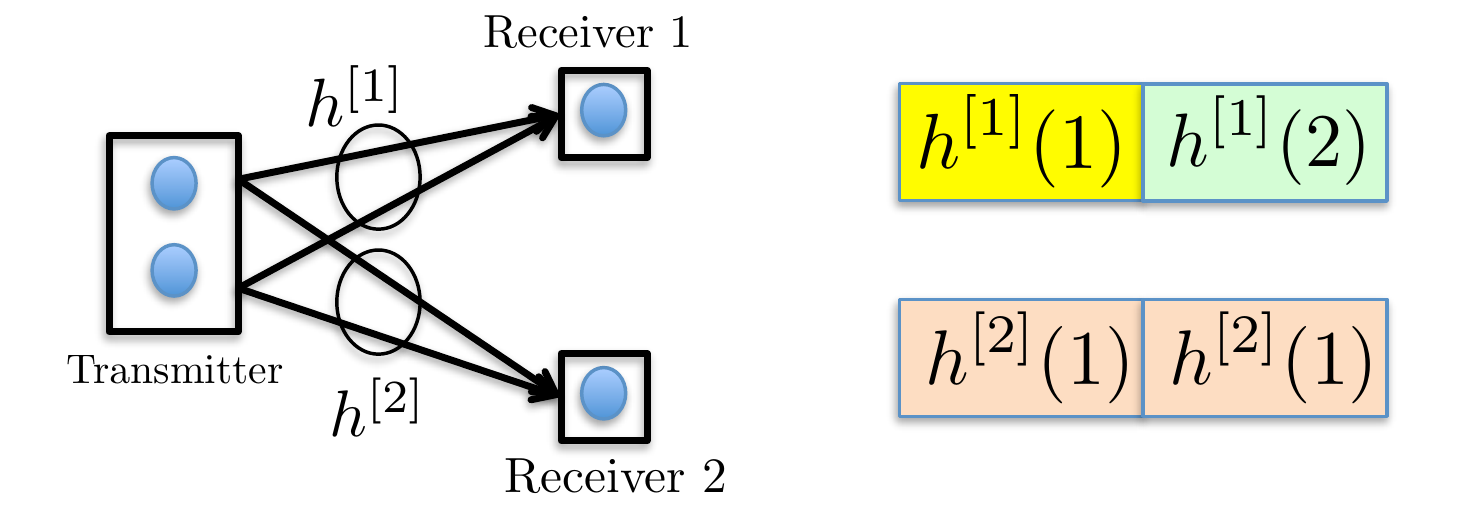}
\caption{2 User MISO BC and the Channel Coherence Structure over a Supersymbol  }
\label{fig:MISOBC}
\end{figure}
Consider the MISO BC, as shown in Figure \ref{fig:MISOBC}, with two antennas at the transmitter and two users with single antenna each, under a staggered block fading model with coherence time $T=2$ and generic channel states. The channel state of user 1, $h^{[1]}=[h^{[1]}_1 ~~h^{[1]}_2]$, is known perfectly to the transmitter while the channel state of user 2, defined similarly, is unknown to the transmitter. Perfect CSIR is assumed at both receivers. 
We do not distinguish between real or complex settings here, as the same arguments apply in either case. The main result for this model is stated in the following theorem.

\begin{theorem}\label{theorem:BConesided}
For the 2 user MISO BC as defined in this section a total of $\frac{3}{2}$ DoF are achievable, almost surely.
\end{theorem}
Note that $\frac{3}{2}$ is the DoF outer bound found for the corresponding scenario in the finite state compound network setting in \cite{Weingarten_Shamai_Kramer}.
A constructive proof is presented next.\\
\proof
We define a supersymbol as comprising of two symbols in the manner shown in Figure \ref{fig:MISOBC}. Thus, user 1's channel state changes over the supersymbol while user 2's channel state maintains a fixed value over a supersymbol.

The received signals of the two users are expressed as follows.
\begin{eqnarray}
\left[\begin{array}{c}y^{[1]}(1)\\y^{[1]}(2)\end{array}\right]&=&\left[\begin{array}{cccc}
h^{[1]}_1(1) & h^{[1]}_2(1)&0&0\\
0&0&h^{[1]}_1(2) & h^{[1]}_2(2)
\end{array}\right]\left[\begin{array}{c}x_1(1)\\x_2(1)\\x_1(2)\\x_2(2)\end{array}\right]+\left[\begin{array}{c}z^{[1]}(1)\\z^{[1]}(2)\end{array}\right]\\
\left[\begin{array}{c}y^{[2]}(1)\\y^{[2]}(2)\end{array}\right]&=&\left[\begin{array}{cccc}
h^{[2]}_1(1) & h^{[2]}_2(1)&0&0\\
0&0&h^{[2]}_1(1) & h^{[2]}_2(1)
\end{array}\right]\left[\begin{array}{c}x_1(1)\\x_2(1)\\x_1(2)\\x_2(2)\end{array}\right]+\left[\begin{array}{c}z^{[2]}(1)\\z^{[2]}(2)\end{array}\right]
\end{eqnarray}
The symbol $y$ is used for received signals, $h$ for channel coefficients, $x$ for input signals and $z$ for additive white Gaussian noise (AWGN). The superscript within square parantheses indicates the user index, the subscript is the antenna index and the index within the round parantheses is the time index within the supersymbol. In compact notation we write equivalently:
\begin{eqnarray}
{\bf Y}^{[1]}&=&{\bf H}^{[1]}{\bf X}+{\bf Z}^{[1]}\\
{\bf Y}^{[2]}&=&{\bf H}^{[2]}{\bf X}+{\bf Z}^{[2]}
\end{eqnarray}

Our goal is to send 2 DoF to user 1 and 1 DoF to user 2, for a total of 3 DoF. Since this is accomplished over two symbols, the normalized total DoF  value is $\frac{3}{2}$. The achievable scheme is based on simple linear beamforming. The transmitted signal is constructed as:
\begin{eqnarray}
{\bf X}=\left[\begin{array}{cc}
1&0\\0&1\\1&0\\0&1
\end{array}\right]\left[\begin{array}{c}u_1^{[1]}\\u_2^{[1]}\end{array}\right]+\left[\begin{array}{c}
h^{[1]}_2(1)\\-h^{[1]}_1(1)\\0\\0
\end{array}\right]u^{[2]}
\end{eqnarray}
Here $u_1^{[1]},u_2^{[1]}$ are the independently encoded scalar Gaussian codewords for user 1, each carrying one DoF, while $u^{[2]}$ is the independently encoded scalar Gaussian codeword for user 2, also carrying one DoF. With this coding scheme, the received signals at user 1 becomes:
\begin{eqnarray}
\left[\begin{array}{c}y^{[1]}(1)\\y^{[1]}(2)\end{array}\right]&=&\left[\begin{array}{cc}
h^{[1]}_1(1) & h^{[1]}_2(1)\\
h^{[1]}_1(2) & h^{[1]}_2(2)
\end{array}\right]\left[\begin{array}{c}u_1^{[1]}\\u_2^{[1]}\end{array}\right]+\left[\begin{array}{c}z^{[1]}(1)\\z^{[1]}(2)\end{array}\right]
\end{eqnarray}
Thus user 1 sees no interference from user 2, and accesses a full rank $2\times 2$ MIMO channel through which he is able to achieve 2 DoF. Now consider the received signal of user 2.
\begin{eqnarray}
\left[\begin{array}{c}y^{[2]}(1)\\y^{[2]}(2)\end{array}\right]&=&\left[\begin{array}{cc}
h^{[2]}_1(1) & h^{[2]}_2(1)\\
h^{[2]}_1 (1)& h^{[2]}_2(1)
\end{array}\right]\left[\begin{array}{c}u_1^{[1]}\\u_2^{[1]}\end{array}\right]+\left[\begin{array}{c}\alpha\\0\end{array}\right]u^{[2]}+\left[\begin{array}{c}z^{[2]}(1)\\z^{[2]}(2)\end{array}\right]\\
&=&\left[\begin{array}{c}1\\1\end{array}\right](h^{[2]}_1(1)u^{[1]}_1+h^{[2]}_2(1)u^{[1]}_2)+\left[\begin{array}{c}\alpha\\0\end{array}\right]u^{[2]}+\left[\begin{array}{c}z^{[2]}(1)\\z^{[2]}(2)\end{array}\right]
\end{eqnarray}
where $\alpha=h^{[2]}_1(1)h^{[1]}_2(1)-h^{[2]}_2(1)h^{[1]}_1(1)\neq 0$, almost surely. Thus, the two streams carrying user 1's signal align into one dimension at user 2's receiver, leaving the remaining dimension to achieve 1 DoF for his desired signal. In this case, a simple projection of the received signal along the vector $[1~~-1]$ provides the interference free signal needed to achieve the desired 1 DoF.
\begin{eqnarray}
y^{[2]}&=&y^{[2]}(1)-y^{[2]}(2)=\alpha u^{[2]}+z^{[2]}(1)-z^{[2]}(2)
\end{eqnarray}
Thus, user 1 achieves 2 DoF and user 2 achieves 1 DoF as desired. Note that the transmitter does not know user 2's channel coefficients at all. Moreover, even for user 1, whose channel is known to the transmitter perfectly, note that only the knowledge of the channel coefficient values over the first coherence interval is used. In other words, the transmitter does not need to know even user 1's channel coefficients over the second coherence interval.

\section{MISO BC with no CSIT for Both Users}
Consider the same 2 user MISO BC channel as in the previous section, with one exception --- now we assume that the channel states of \emph{both} users are unknown to the transmitter. 
\begin{figure}[!t]
\centering
\includegraphics[width=5in]{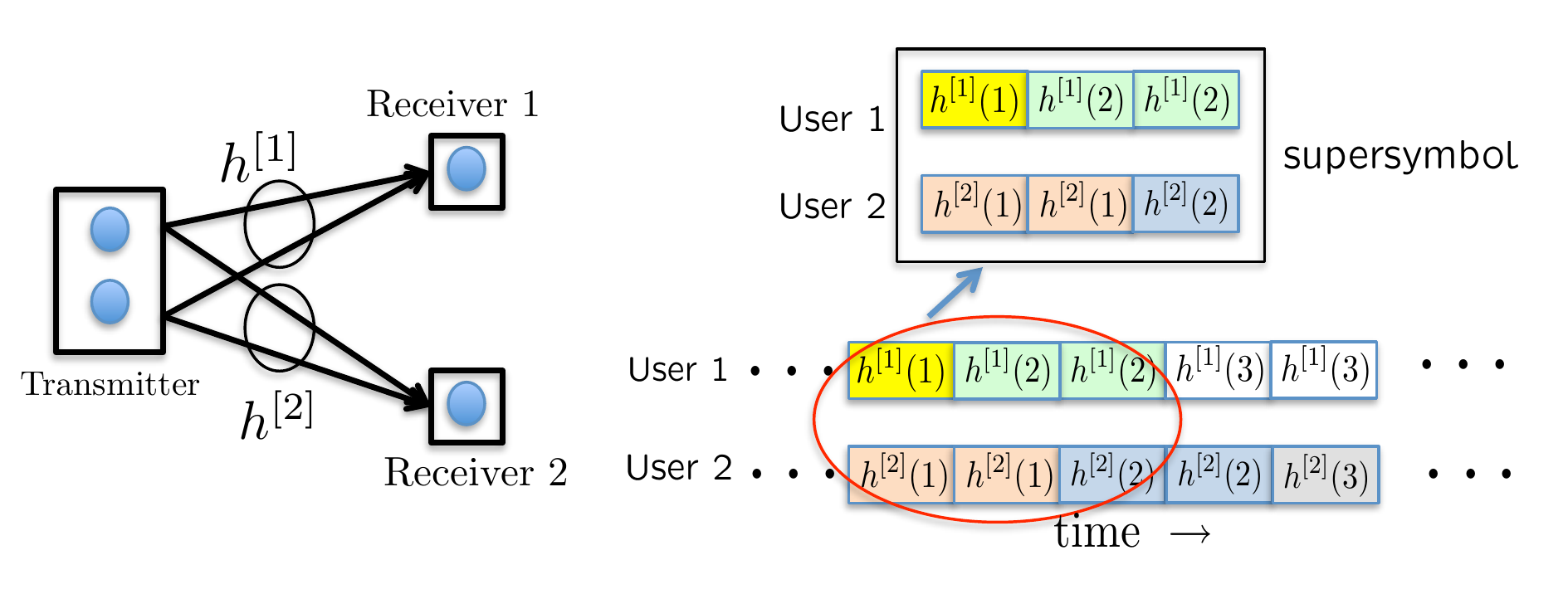}
\caption{Supersymbol Structure for Theorem \ref{theorem:BCtwosided}}
\label{fig:bc3staggered}
\end{figure}

The main result for this model is stated in the following theorem.

\begin{theorem}\label{theorem:BCtwosided}
For the 2 user MISO BC as defined in this section a total of $\frac{4}{3}$ DoF are achievable, almost surely.
\end{theorem}
Note that $\frac{4}{3}$ is the DoF outer bound found for the corresponding scenario in the finite state compound network setting in \cite{Weingarten_Shamai_Kramer}.
A constructive proof is presented next.\\
\proof
While our coherence model is the same as the previous section, i.e. staggered block fading with coherence time $T=2$ for both users, in this case we define the supersymbol as  comprised of three symbols. As shown in Figure \ref{fig:bc3staggered} the 3 symbols are chosen such that the channel state of user 1  changes after the first symbol and remains fixed for the last 2 symbols, while the channel state of user 2 is fixed for the first 2 symbols and changes in the last symbol. The received signals over one supersymbol defined in this manner, are expressed as follows.
\begin{eqnarray*}
\left[\begin{array}{c}y^{[1]}(1)\\y^{[1]}(2)\\y^{[1]}(3)\end{array}\right]&=&\left[\begin{array}{cccccc}
h^{[1]}_{1}(1) & h^{[1]}_{2}(1)&0&0&0&0\\
0&0&h^{[1]}_{1}(2) & h^{[1]}_{2}(2)&0&0\\
0&0&0&0&h^{[1]}_{1}(2) & h^{[1]}_{2}(2)
\end{array}\right]\left[\begin{array}{c}x_1(1)\\x_2(1)\\x_1(2)\\x_2(2)\\x_1(3)\\x_2(3) \end{array}\right]+\left[\begin{array}{c}z^{[1]}(1)\\z^{[1]}(2)\\z^{[1]}(3) \end{array}\right]\nonumber\\
\left[\begin{array}{c}y^{[2]}(1)\\y^{[2]}(2)\\y^{[2]}(3)\end{array}\right]&=&\left[\begin{array}{cccccc}
h^{[2]}_{1}(1) & h^{[2]}_{2}(1)&0&0&0&0\\
0&0&h^{[2]}_{1}(1) & h^{[2]}_{2}(1)&0&0\\
0&0&0&0&h^{[2]}_{1}(2) & h^{[2]}_{2}(2)
\end{array}\right]\left[\begin{array}{c}x_1(1)\\x_2(1)\\x_1(2)\\x_2(2)\\x_1(3)\\x_2(3) \end{array}\right]+\left[\begin{array}{c}z^{[2]}(1)\\z^{[2]}(2)\\z^{[2]}(3) \end{array}\right]
\end{eqnarray*}

Our goal is to achieve two DoF for each user over this supersymbol consisting of $3$ symbols, to achieve an overall normalized DoF equal to $\frac{4}{3}$, consistent with the outer bounds in \cite{Lapidoth_Shamai_Wigger_BC,Weingarten_Shamai_Kramer} for similar settings. To accomplish this objective, we construct the input vector ${\bf X}$ as follows.
\begin{eqnarray}
{\bf X}=\left[\begin{array}{cc}
1&0\\
0&1\\
1&0\\
0&1\\
0&0\\
0&0
\end{array}\right]\left[\begin{array}{c}u_1^{[1]}\\u_2^{[1]}\end{array}\right]+\left[\begin{array}{cc}
0&0\\
0&0\\
1&0\\
0&1\\
1&0\\
0&1
\end{array}\right]\left[\begin{array}{c}u_1^{[2]}\\u_2^{[2]}\end{array}\right]
\end{eqnarray}
where $u_1^{[k]}, u_2^{[k]}$ are the independently encoded scalar Gaussian codeword symbols for user $k=1,2$ respectively, each carrying one DoF. Note that the beamforming vectors do not depend on the values of channel coefficients. With this scheme the signal at receiver 1 becomes:
\begin{eqnarray*}
\left[\begin{array}{c}y^{[1]}(1)\\y^{[1]}(2)\\y^{[1]}(3)\end{array}\right]&=&
\underbrace{\left[\begin{array}{cc}
h^{[1]}_{1}(1) & h^{[1]}_{2}(1)\\
h^{[1]}_{1}(2) & h^{[1]}_{2}(2)\\
0&0
\end{array}\right]}_{\mbox{rank}=2}\left[\begin{array}{c}u_1^{[1]}\\u_2^{[1]}\end{array}\right]
+
\underbrace{\left[\begin{array}{cc}
0&0\\
h^{[1]}_{1}(2) & h^{[1]}_{2}(2)\\
h^{[1]}_{1}(2) & h^{[1]}_{2}(2)
\end{array}\right]}_{\mbox{rank}=1}\left[\begin{array}{c}u_1^{[2]}\\u_2^{[2]}\end{array}\right]+\left[\begin{array}{c}z^{[1]}(1)\\z^{[1]}(2)\\z^{[1]}(3) \end{array}\right]\\
&=&\left[\begin{array}{cc}
h^{[1]}_{1}(1) & h^{[1]}_{2}(1)\\
h^{[1]}_{1}(2) & h^{[1]}_{2}(2)\\
0&0
\end{array}\right]\left[\begin{array}{c}u_1^{[1]}\\u_2^{[1]}\end{array}\right]
+
\left[\begin{array}{c}
0\\1\\1
\end{array}\right](h^{[1]}_{1}(2)u_1^{[2]}+h^{[1]}_{2}(2)u_2^{[2]})+\left[\begin{array}{c}z^{[1]}(1)\\z^{[1]}(2)\\z^{[1]}(3) \end{array}\right]
\end{eqnarray*}
Thus in the $3$ dimensional received signal space of receiver $1$, the interference from user 2's signal aligns within 1 dimension, while the desired signals, carrying 2 DoF, occupy two linear independent dimensions. It remains to check that the two desired signal dimensions do not overlap with the one interference dimension. This is easily verified as the determinant of the matrix
\begin{eqnarray}
\left[\begin{array}{ccc}
h^{[1]}_{1}(1) & h^{[1]}_{2}(1)&0\\
h^{[1]}_{1}(2) & h^{[1]}_{2}(2)&1\\
0&0&1
\end{array}\right]
\end{eqnarray}
is not equal to zero almost surely. In other words, user 1 is able to achieve $2$ DoF. By symmetry the same arguments can be used to show that user 2 achieves $2$ DoF as well, so that the desired $\frac{4}{3}$ (normalized) DoF are achieved.

\section{The X Channel}
\begin{figure}[!h]
\centering
\includegraphics[width=5in]{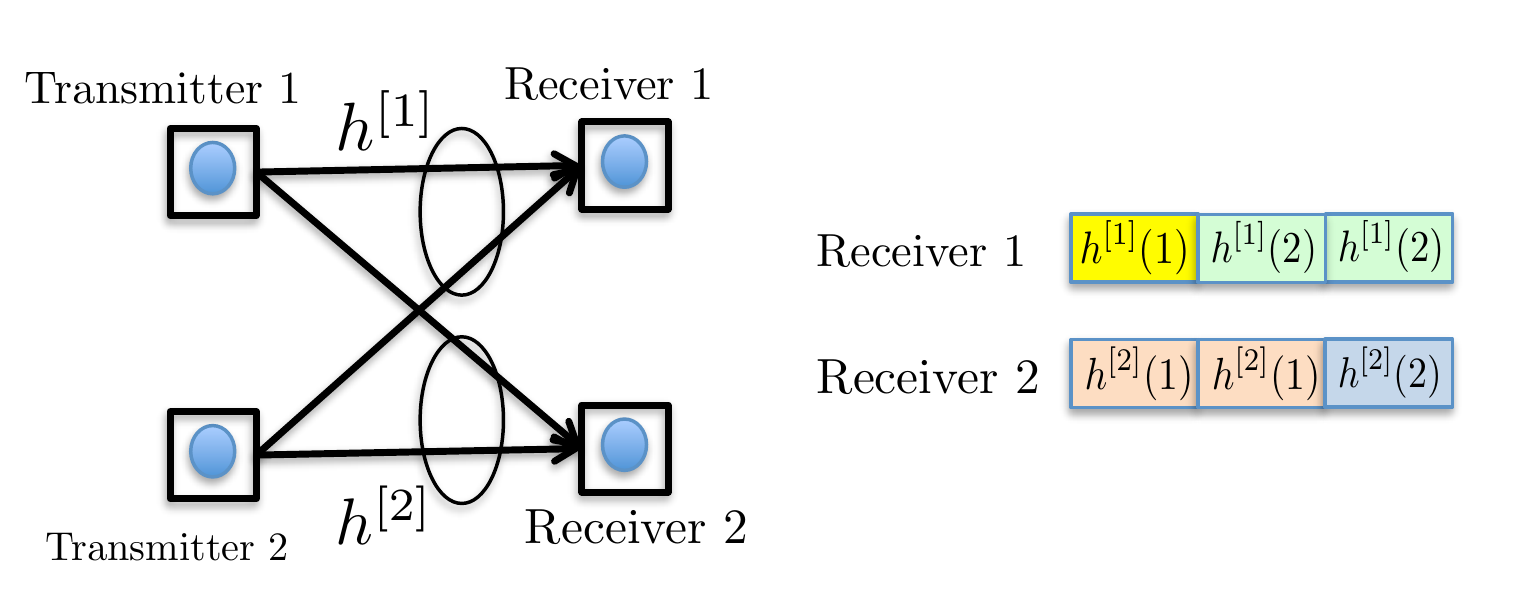}
\caption{X Channel and the Supersymbol Structure}
\label{fig:X}
\end{figure}
The X channel that we consider in this section, consists of two transmitters and two receivers, each equipped with a single antenna and 4 independent messages, one for each transmitter-receiver pair. Note that if we separate the two antennas at the transmitter of the MISO BC considered in the previous section to form two separate transmitters, i.e., if we do not allow joint processing of signals or common knowledge of messages at the two transmit antennas of the MISO BC, then we obtain the X channel. The remaining assumptions -- no CSIT, perfect CSIR -- and staggered block fading with coherence time $T=2$, are the same as the MISO BC in the previous section. The channel input output relationships are also the same as the previous section. The DoF result for the X channel is stated in the following theorem.

\begin{theorem}\label{theorem:X}
For the X channel as defined in this section a total of $\frac{4}{3}$ DoF are achievable, almost surely.
\end{theorem}
Note that $\frac{4}{3}$ is the DoF outer bound found even with perfect CSIT. Therefore it is also an outer bound with no CSIT.\\
\proof
The proof of Theorem \ref{theorem:X} follows trivially from the proof of Theorem \ref{theorem:BCtwosided}. Note that no cooperation between the two transmit antennas is needed for the achievable scheme of the 2 user MISO BC with no CSIT described in the previous section. Thus, the same achievable scheme can be applied directly for the 2 user X channel as well. In both cases $\frac{4}{3}$ DoF are achieved. Note that this insight is consistent with the result from the finite state compound setting found in \cite{Gou_Jafar_Wang}, where also it is found that with enough channel uncertainty, the MISO BC devolves into the X channel as the DoF benefits of joint processing across transmit antennas are lost.

\section{The 2 User MIMO Interference Channel}
\begin{figure}[!h]
\centering
\includegraphics[width=5in]{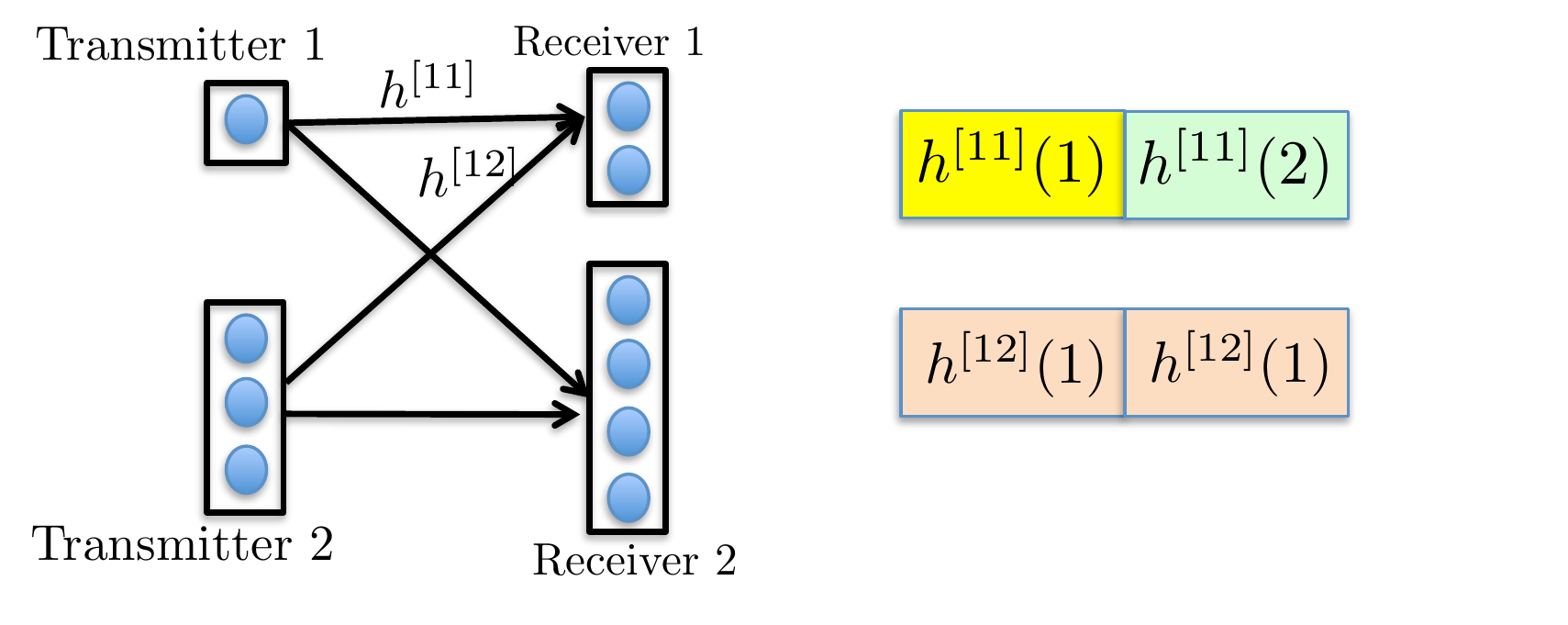}
\caption{MIMO Interference Channel and the Supersymbol Structure}
\label{fig:MIMOIC}
\end{figure}

This example is based on the open problem in \cite{Huang_Jafar_Shamai_Vishwanath}. We have a two user MIMO interference channel with $1$ and $3$ antennas at the transmitters and $2$ and $4$ antennas at their corresponding receivers, respectively. As in all preceding sections we assume no CSIT, perfect CSIR, and a staggered block fading model with coherence time $T=2$, with one subtle difference. The coherence times are staggered as seen by the receiver, rather than the transmitter, i.e., from Receiver 1's perspective, the block fading channels from transmitter 1 and 2 are staggered. For the problem we are interested in, it does not matter what temporal correlation is seen by Receiver 2. As seen by Receiver 2, the channels may have any coherence time, and the fading blocks may be aligned or staggered. Since the key to this problem is the possibility of interference alignment only at Receiver 1 --- Receiver 2 has enough antennas to separate all signal and interference --- it is not surprising that only the fading block structure at Receiver 1 is of significance. 

 The question posed in \cite{Huang_Jafar_Shamai_Vishwanath} is the following -- what is the maximum DoF achievable by user 2 simultaneously as user 1 achieves his maximum (one) DoF ? The best outer bound found in \cite{Huang_Jafar_Shamai_Vishwanath} is $\frac{3}{2}$ but the best inner bound is only able to achieve $1$ DoF for user 2. 
 
 The DoF result for this channel is presented in the following theorem.
\begin{theorem}\label{theorem:MIMOIC}
For the 2 user MIMO interference channel defined in this section, users 1 and 2 can simultaneously achieve $1$ and $\frac{3}{2}$ DoF, respectively, almost surely.
\end{theorem}
Note that this result also matches the outer bound found in \cite{Huang_Jafar_Shamai_Vishwanath}. The key to this achievability, as suggested in \cite{Huang_Jafar_Shamai_Vishwanath} is interference alignment at Receiver 1. A constructive proof follows next.\\
\proof Our goal is for user 1 to achieve $2$ DoF and for user $2$ to achieve $3$ DoF over this two symbol extension, which corresponds to normalized values of $1$ and $\frac{3}{2}$, respectively. Interference alignment is needed to accomplish this objective and like all previous examples we rely on linear beamforming techniques.

 For this example, we define a supersymbol as comprised of two symbols, which leads to the structure shown in Figure \ref{fig:MIMOIC}.  Within a supersymbol, the signal at Receiver 1 is expressed as:
\begin{eqnarray}
\left[\begin{array}{c}y^{[1]}_1(1)\\y^{[1]}_2(1)\\y^{[1]}_1(2)\\y^{[1]}_2(2)\end{array}\right]&=&\left[\begin{array}{cccccc}
h^{[12]}_{11}(1) & h^{[12]}_{12}(1)&h^{[12]}_{13}(1)&0&0&0\\
h^{[12]}_{21} (1)& h^{[12]}_{22}(1)&h^{[12]}_{23}(1)&0&0&0\\
0&0&0&h^{[12]}_{11} (1)& h^{[12]}_{12}(1)&h^{[12]}_{13}(1)\\
0&0&0&h^{[12]}_{21}(1) & h^{[12]}_{22}(1)&h^{[12]}_{23}(1)
\end{array}\right]\left[\begin{array}{c}x^{[2]}_1(1)\\x^{[2]}_2(1)\\x^{[2]}_3(1)\\x^{[2]}_1(2)\\x^{[2]}_2(2)\\x^{[2]}_3(2) \end{array}\right]\nonumber\\
&&+
\left[\begin{array}{cc}
h^{[11]}_{11}(1)&0\\
h^{[11]}_{21}(1)&0\\
0&h^{[11]}_{11}(2)\\
0&h^{[11]}_{21}(2)
\end{array}\right]\left[\begin{array}{c}x^{[1]}(1)\\x^{[1]}(2)\end{array}\right]
+\left[\begin{array}{c}z^{[1]}_1(1)\\z^{[1]}_2(1)\\z^{[1]}_1(2)\\z^{[1]}_2(2)\end{array}\right]
\end{eqnarray}
Equivalently, in compact notation
\begin{eqnarray}
{\bf Y}^{[1]}&=&{\bf H}^{[12}{\bf X}^{[2]}+{\bf H}^{[11]}{\bf X}^{[1]}+{\bf Z}^{[1]}
\end{eqnarray}
The key to the alignment is to design user 2's signal as follows.
\begin{eqnarray}
{\bf X}^{[2]}=\left[\begin{array}{ccc}
1&0&0\\0&1&0\\0&0&1\\
1&0&0\\0&1&0\\0&0&1
\end{array}\right]\left[\begin{array}{c}u^{[2]}_1\\u^{[2]}_2\\u^{[2]}_3\end{array}\right]
\end{eqnarray}
where $u^{[1]}_i$ is the $i^{th}$ independently coded scalar stream sent from transmitter 2. Each stream carries 1 DoF. With this coding scheme, the signal at Receiver 1 becomes
\begin{eqnarray*}
{\bf Y}^{[1]}&=&\underbrace{\left[\begin{array}{ccc}
h^{[12]}_{11}(1) & h^{[12]}_{12}(1)&h^{[12]}_{13}(1)\\
h^{[12]}_{21}(1) & h^{[12]}_{22}(1)&h^{[12]}_{23}(1)\\
h^{[12]}_{11}(1) & h^{[12]}_{12}(1)&h^{[12]}_{13}(1)\\
h^{[12]}_{21}(1) & h^{[12]}_{22}(1)&h^{[12]}_{23}(1)
\end{array}\right]}_{\mbox{rank}=2}
\left[\begin{array}{c}u^{[2]}_1\\u^{[2]}_2\\u^{[2]}_3\end{array}\right]
+
\left[\begin{array}{cc}
h^{[11]}_{11}(1)&0\\
h^{[11]}_{21}(1)&0\\
0&h^{[11]}_{11}(2)\\
0&h^{[11]}_{21}(2)
\end{array}\right]\left[\begin{array}{c}x^{[1]}(1)\\x^{[1]}(2)\end{array}\right]
+\left[\begin{array}{c}z^{[1]}_1(1)\\z^{[1]}_2(1)\\z^{[1]}_1(2)\\z^{[1]}_2(2)\end{array}\right]
\end{eqnarray*}
The interference alignment is manifested in the rank deficiency of the $4\times 3$ effective channel matrix between transmitter $1$ and receiver $2$. Note that the first and third rows are identical, as are the second and fourth rows. Thus this matrix has rank only $2$. Equivalently, the interference space seen by Receiver 1 is spanned by the first two columns of this matrix. In a $4$ dimensional received space at Receiver 1, since interference spans only two dimensions, the remaining $2$ dimensions are available to achieve its desired $2$ DoF. However, we must ensure that the desired signals arrive along linearly independent directions from the interference. In other words, the following matrix must be full rank.
\begin{eqnarray}
\left[\begin{array}{cccc}
h^{[11]}_{11}(1)&0&h^{[12]}_{11} & h^{[12]}_{12}\\
h^{[11]}_{21}(1)&0&h^{[12]}_{21} & h^{[12]}_{22}\\
0&h^{[11]}_{11}(2)&h^{[12]}_{11} & h^{[12]}_{12}\\
0&h^{[11]}_{21}(2)&h^{[12]}_{21} & h^{[12]}_{22}
\end{array}\right]
\end{eqnarray}
Here the first two columns span the desired signal space while the last two columns span the interference space. It is easy to compute the determinant of this matrix, which is seen to be non-zero as long as $(h^{[11]}_{11}(1), h^{[11]}_{21}(1))\neq (h^{[11]}_{11}(2), h^{[11]}_{21}(2))$, i.e. the channel $h^{[11]}$ changes from one coherence block to another. Since this is true almost surely (by definition of the block fading model),  user 1 is able to achieve his maximum $2$ DoF over $2$ symbols. The achievability of user 2's three DoF is straightforward, because with $4$ receive antennas, receiver 2 is able to invert the channel from both transmitters simultaneously, which allows it to separate the two users' signals, regardless of the coherence times.
\section{The K User  Interference Channel}
\begin{figure}[!h]
\centering
\includegraphics[width=6in]{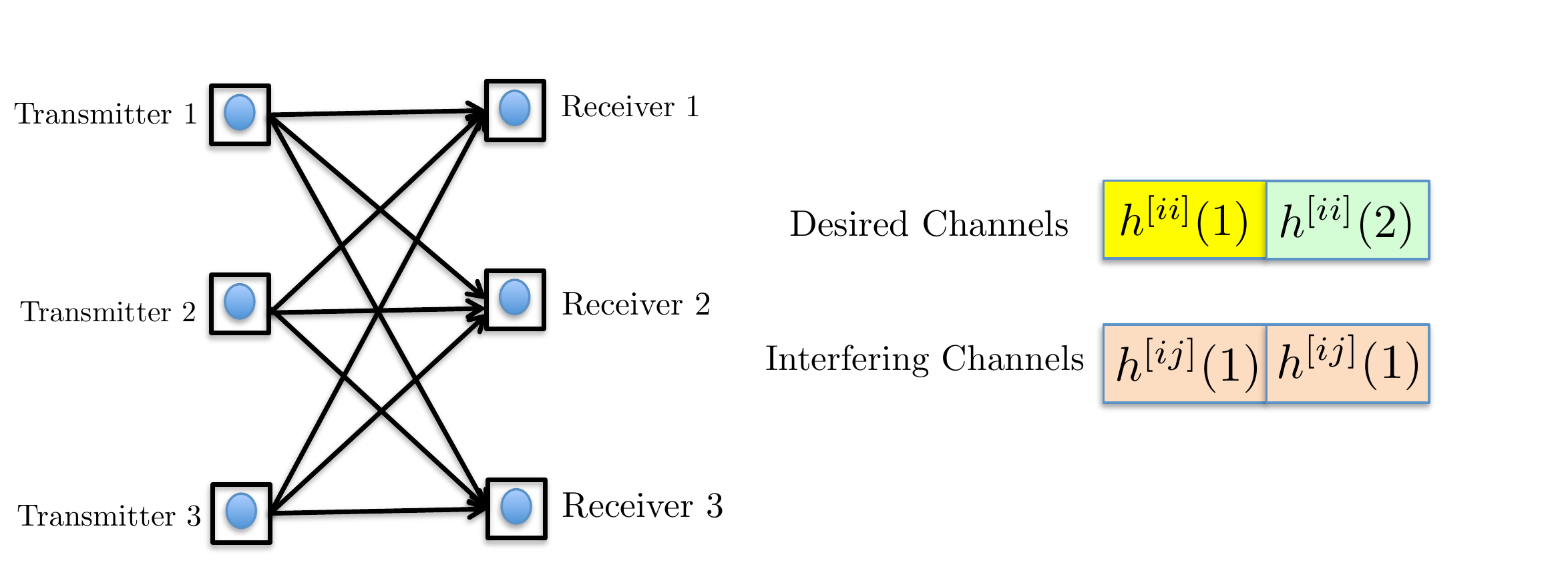}
\caption{3 User Interference Channel and the Supersymbol Structure}
\label{fig:3userIC}
\end{figure}The final interference alignment problem that we consider in this paper is the $K$ user interference channel, comprised of $K$ transmitters, $K$ receivers, each equipped with a single antenna, and $K$ independent messages, one from each transmitter to its corresponding receiver. As usual we assume no CSIT, perfect CSIR and staggered block fading with coherence time $T=2$. Specifically, we assume a rather specialized form of channel coherence structure. All the direct channels (carrying desired signals) have the same fading block boundaries. All the cross-channels (carrying interference) also have the same fading block boundaries. However, the fading block boundaries of the desired channels are staggered with respect to the fading block boundaries of the cross-channels. Admittedly this correlation structure is less natural than other cases studied above. Nevertheless it serves to show that interference alignment can be achieved without the knowledge of channel coefficient values in this setting as well. 

 A 3 user example is shown in Figure \ref{fig:3userIC}. The DoF result for this setting is stated in the following theorem. 
\begin{theorem}\label{theorem:KuserIC}
For the K user interference channel defined in this section, a total of $\frac{K}{2}$ DoF are achievable, almost surely.
\end{theorem}
Note that $\frac{K}{2}$ is the DoF outer bound even with perfect CSIT \cite{Nosratinia_Madsen}. Therefore it is also an outer bound with no CSIT. A constructive achievability proof is presented next.\\
\proof For simplicity of exposition we present the proof for the $K=3$ user interference channel. The extension to $K>3$ is straightforward.

We define a supersymbol as comprised of two symbols in the manner shown in Figure \ref{fig:3userIC}, i.e., the desired channels change values but the interfering channels are held constant within a supersymbol. The signal received by Receiver 1 over one supersymbol defined in this manner, is expressed as follows.
\begin{eqnarray*}
\left[\begin{array}{c}y^{[1]}(1)\\y^{[1]}(2)\end{array}\right]&=&\left[\begin{array}{cc}
h^{[11]}(1) & 0\\
0& h^{[11]}(2)
\end{array}\right]\left[\begin{array}{c}x^{[1]}(1)\\x^{[1]}(2)\end{array}\right]
+\left[\begin{array}{cc}
h^{[12]}(1) & 0\\
0& h^{[12]}(1)
\end{array}\right]\left[\begin{array}{c}x^{[2]}(1)\\x^{[2]}(2)\end{array}\right]\nonumber\\
&&+\left[\begin{array}{cc}
h^{[13]}(1) & 0\\
0& h^{[13]}(1)
\end{array}\right]\left[\begin{array}{c}x^{[3]}(1)\\x^{[3]}(2)\end{array}\right]+\left[\begin{array}{c}z^{[1]}(1)\\z^{[1]}(2)\end{array}\right]
\end{eqnarray*}
Each user $k=1,2,3$, sends one scalar coded stream $u^{[k]}$ carrying one DoF along the beamforming vector $[1   ~~~1]^T$. The received signal can then be written as:
\begin{eqnarray*}
\left[\begin{array}{c}y^{[1]}(1)\\y^{[1]}(2)\end{array}\right]&=&\left[\begin{array}{c}
h^{[11]}(1) \\
h^{[11]}(2)
\end{array}\right]u^{[1]}
+\left[\begin{array}{c}
1\\
1
\end{array}\right](h^{[12]}(1)u^{[2]}+h^{[13]}(1)u^{[3]})+\left[\begin{array}{c}z^{[1]}(1)\\z^{[1]}(2)\end{array}\right]
\end{eqnarray*}
Thus, in the two dimensional received signal space of Receiver 1, all the interference aligns along the vector $[1~~~1]^T$ and the desired signal, which is received in a linearly independent direction almost surely, can be separated from interference to yield one DoF. By symmetry, the same is true for each receiver and a total of $K$ DoF are achieved over 2 symbols. In other words, the $K$ user interference channel has $K/2$ DoF almost surely.
\section{Conclusion}
The main contribution of this work is the idea that channel correlations can be exploited to achieve interference alignment, even when the transmitter has no information about the precise values taken by the channel coefficients, which may be drawn from a continuum of values. While we present our results in the time domain setting, for the developments in this paper there is no fundamental distinction between time and frequency dimensions. Hence the channel correlations could be in time or frequency and one can translate easily from time-variations to frequency-selectivity, coherence-time to coherence bandwidth etc. Under a staggered block fading model, what is surprising is not only that interference alignment is achieved without CSIT, but also that the alignment schemes are quite simple. While the present work focuses on some limited examples to convey the fundamental idea, our continuing work aims at generalizing these results to arbitrary temporal correlation models. In particular, the validity of the Lapidoth-Shamai-Wigger conjecture \cite{Lapidoth_Shamai_Wigger_BC} for the MISO BC is an important benchmark for further progress in this direction. Note that in this work our channel model differs from \cite{Lapidoth_Shamai_Wigger_BC} only in the assumption that channel values within a fading block are identical, which --- even with the very small coherence time  $T=2$ needed here  --- makes the differential entropy rate of the channel sequence equal to $-\infty$. Any reasonable departure from the assumption of precisely identical channel coefficients within a coherence block will remove this limitation. However, what remains to be established is if such a departure would necessarily lead to a collapse of all DoF for the MISO BC.

\bibliographystyle{IEEEtran}
\bibliography{Thesis}

\end{document}